\documentclass[aps,pra,twocolumn,superscriptaddress]{revtex4}
\usepackage{amsmath,amsfonts,graphicx}
\usepackage[caption=false]{subfig}

\begin{document}

\title{Simple derivation of the Weyl and Dirac quantum cellular automata\\}
\author{Philippe Raynal}
\affiliation{Centre for Quantum Technologies, National University of Singapore, Singapore}
\affiliation{University Scholars Programme, National University of Singapore, Singapore}
\date{\today}

\begin{abstract}
We consider quantum cellular automata on a body-centred cubic lattice and provide a simple derivation of the only two homogenous, local, isotropic, and unitary two-dimensional automata {[G. M. D'Ariano and P. Perinotti, Physical Review A 90, 062106 (2014)]}. Our derivation relies on the notion of Gram matrix and emphasises the link between the transition matrices that characterise the automata and the body-centred cubic lattice: The transition matrices essentially are the matrix representation of the vertices of the lattice's primitive cell. As expected, the dynamics of these two automata reduce to the Weyl equation in the limit of small wave vectors and continuous time. We also briefly examine the four-dimensional case where we find two one-parameter families of automata that reduce to the Dirac equation in a suitable limit.
\end{abstract}

\maketitle 

\section{Introduction}
In the forties, Von Neumann proposed a discrete model of computation where finite-state machines arranged in a grid, locally process information in discrete time-steps \cite{vonneumann1966}. Von Neumann's motivation to introduce cellular automata was to obtain a complex behaviour from an array of simple processing units. Later, Feynman suggested applying the laws of quantum mechanics to computation \cite{feynman1982,feynman1986}. This idea naturally led to the formalisation of quantum computation \cite{benioff1980,deutsch1985} in general and Quantum Cellular Automata (QCA) in particular \cite{lloyd1993,watrous1995,werner2004}. In parallel to QCA, other models of quantum computation have been developed like circuit-based quantum computation \cite{deutsch1989}, measurement-based quantum computation \cite{raussendorf2001}, or adiabatic quantum computation \cite{farhi2001}.

Around the same time, conceptual links between physics and computer science were examined. A famous example is the relation between logical reversibility and physical reversibility investigated by Landauer \cite{landauer1961} and culminating with the development of reversible computation \cite{bennett1973,fredkin1982}. A more recent instance, motivated by the believe that spacetime should ultimately be discrete, concerns the emergence of (continuous) dynamical laws of physics from (discrete) computation. This question was first discussed by Suze \cite{suze1969} and later by Weehler \cite{wheeler1985}. The first significant step towards emergent physical laws came from Bialynicki-Birula who considered a two-dimensional QCA on a body-centred cubic lattice \cite{birula1994}. Bialynicki-Birula provided an example of a homogenous, linear, and unitary QCA and showed that the dynamic of any QCA that fulfils these constraints necessarily reduces to the Weyl equation in the limit of small wave vectors and continuous time. He also showed that the simple cubic lattice admits no homogenous, linear, and unitary QCA. In passing, Bialynicki-Birula proposed a four-dimensional QCA that reduces to the Dirac equation in the limit of small wave vectors and continuous time. A couple of years latter, Meyer showed that there is no nontrivial homogeneous, local and unitary  one-dimensional cellular automaton on a unidimensional lattice \cite{meyer1996}, a result later generalised to higher dimensions in \cite{meyer1996_2} and \cite{dariano2016_2}. Recently, D'Ariano {\it et al.} extended Bialynicki-Birula's and Meyer's no-go results to the rhombohedral lattice. In addition, they showed that there are not one but two homogeneous, local, linear, isotropic, and unitary QCA for the body-centred cubic lattice, up to unitary conjugation \cite{dariano2014}. However, the derivation is rather lengthy and technical. In the present paper, we provide a short and simple derivation of these two unitarily-inequivalent automata. Our derivation uses the notion of Gram matrix and emphasises the relation between the transition matrices that describe the unitary evolution of the automata and the body-centred cubic lattice: The transition matrices essentially are the matrix representation of the vertices of the lattice's primitive cell, here a central vertex surrounded by a tetrahedron and its dual. Although we only consider the body-centred cubic lattice in this paper, our technique can directly be used for all Bravais lattices. Let us add here that the Klein-Gordon equation \cite{arrighi2013}, Maxwell's equations \cite{dariano2015}, Lorentz invariance \cite{arrighi2014,farrelly2014,dariano2016} and curved spacetime \cite{molfetta2013,arrighi2016} have also been investigated in the framework of QCA.

This paper is organised as follows. In Section II, we recall the general framework to discuss QCA on an abstract lattice as established in \cite{dariano2014}. In Section III, we focus our attention on two-dimensional homogenous, local, isotropic, and unitary automata on a body-centred cubic lattice and derive the only two unitarity-inequivalent solutions. This short and simple derivation is the main result of our paper. We briefly examine the four-dimensional case in Section IV and find two one-parameter families of automata without completely solving the case. These automata reduce to the Dirac equation in a suitable limit. We conclude in Section V.

\section{Quantum cellular automata}

We follow the  general framework developed in \cite{dariano2014} to describe a homogenous, local, isotropic, and unitary automaton on an abstract lattice. A QCA is defined as a numerable set $\cal G$ of identical cells that evolve in identical discrete time-steps. Each cell is a quantum system and the discrete evolution is unitary. We further focus our attention on the case of fermionic QCA, where the quantum system of each cell is described by a finite number of fermionic field operators satisfying the usual anti-commutation relations \cite{dariano2017} (See \cite{werner2004} for a formal definition of a QCA in the qudit case.). We will later restrict the abstract set $\cal G$ to the abelian group of translations in a three-dimensional body-centred cubic lattice. The assumption that the cells and the time-steps are all identical represents our homogeneity assumption.

\begin{figure}[ht!]
{\includegraphics[scale=.22]{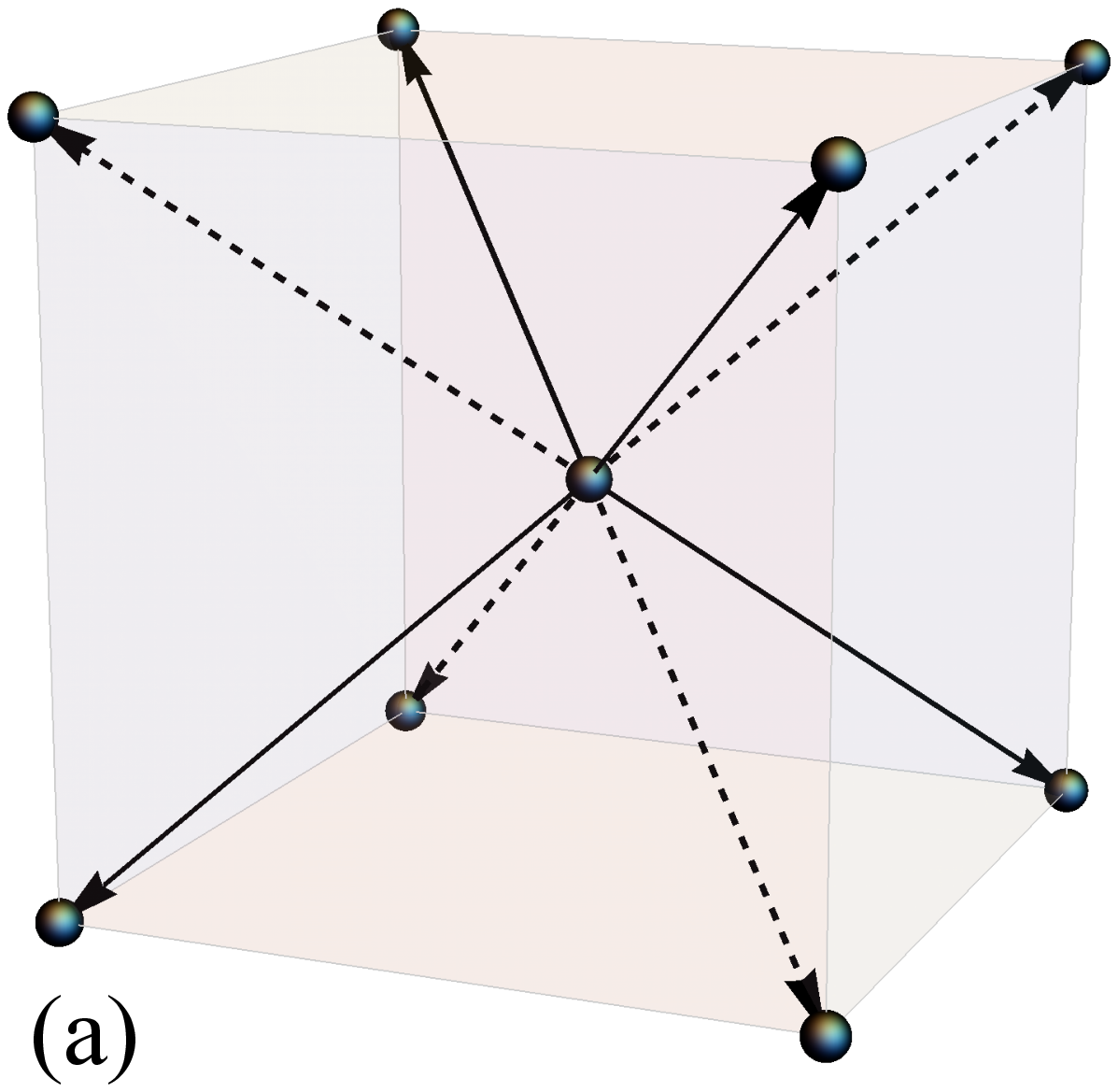}}
{\includegraphics[scale=.22]{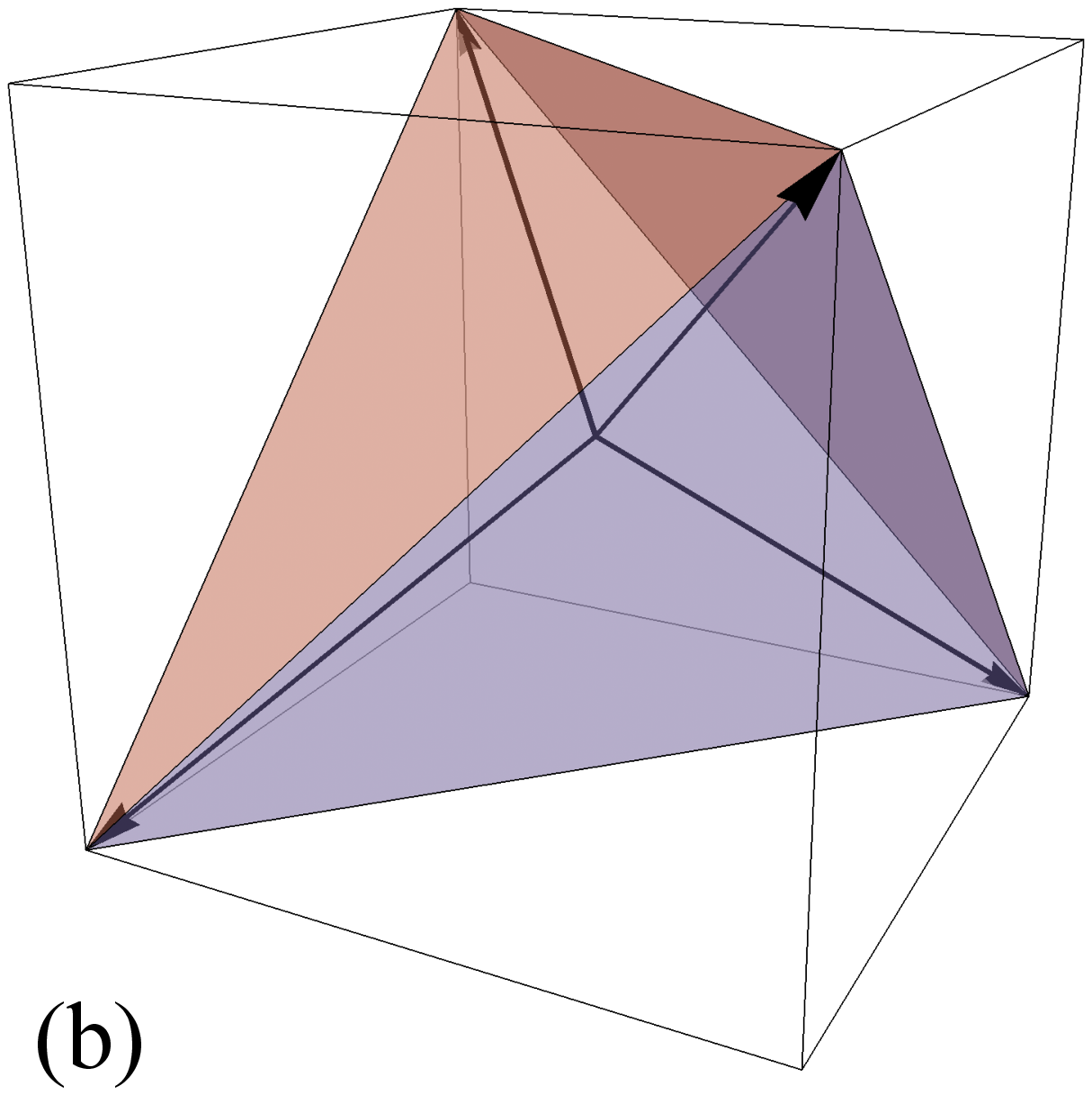}}
{\includegraphics[scale=.22]{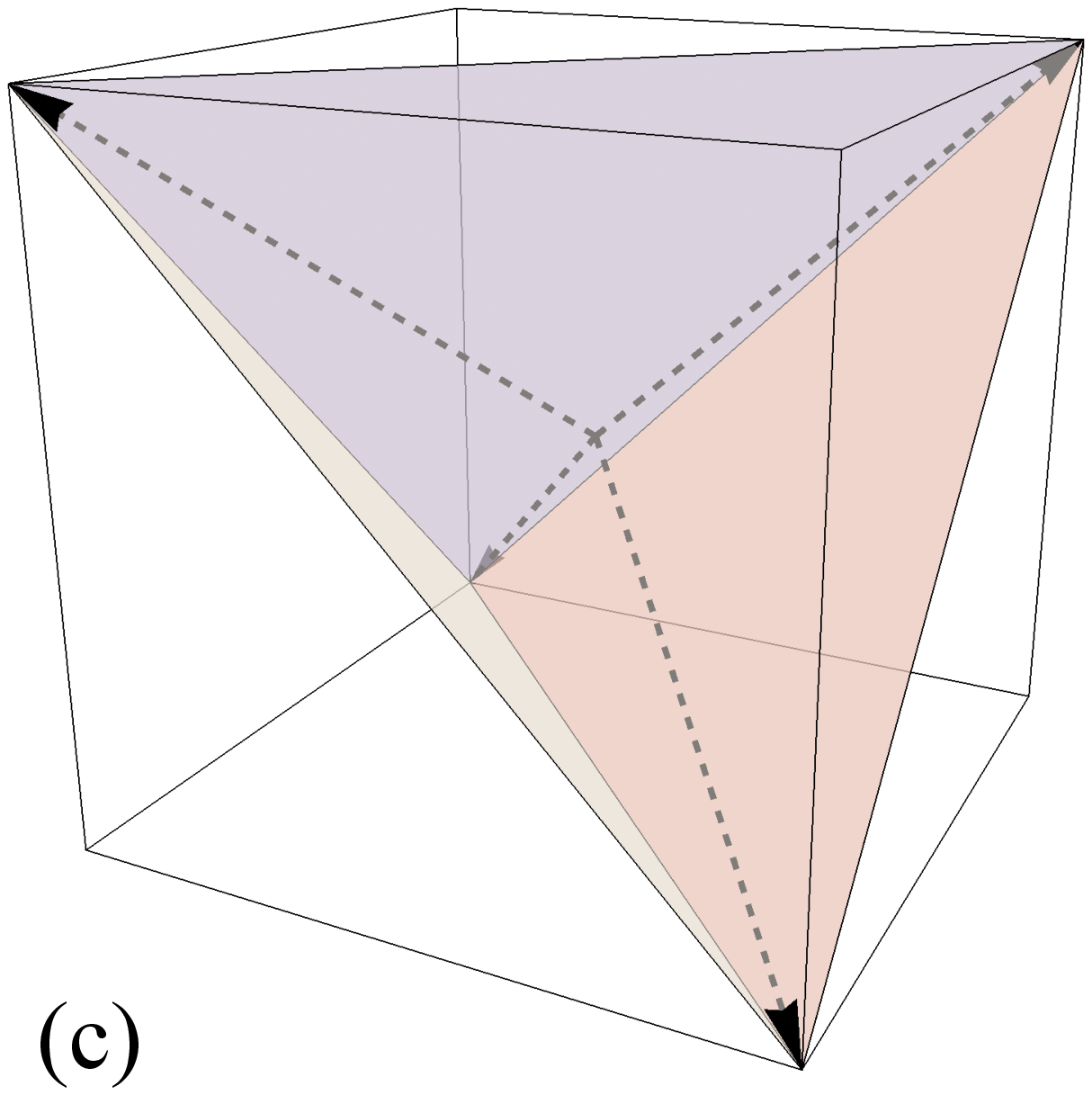}}
\captionsetup{justification=justified}\caption{The primitive cell of the body-centred cubic lattice can be seen as a central vertex surrounded by a tetrahedron and its dual. (a) Graphical representation of the primitive cell of the body-centred cubic lattice with a central vertex and its eight neighbouring vertices (thick and dashed vectors). (b) Graphical representation of the tetrahedron generated by the four vertices in $S_+$ (thick vectors). (c) Graphical representation of the dual tetrahedron generated by the four vertices in $S_-$ (dashed vectors).}
\label{fig:tetra}
\end{figure}

Next, we define the finite set $V_g$ of neighbouring cells of a given cell $g$ and we assume the deterministic evolution of the automaton to be local, that is, each cell $g$ only interacts with a finite number $|V_g|$ of neighbouring cells within a single step of computation.
We now have a QCA on a graph $\Gamma(V,E)$ with vertex set $V=\{g \in \cal G\}$ and edge set $E=\{(g,g')|g\in {\cal G}, g' \in V_g\}$ where a vector $\psi_g$ is associated to each vertex and a transition function $A_{g,g'}$ is associated to each directed edge $(g,g')$. The components of the vector $\psi_g$ are the field operators associated to the quantum system at site $g$. Furthermore, the homogeneity assumption means that all vertices are identical so that (1) the transition matrices $A_{g,g'}$ are independent of the vertex $g$ and (2) each cell has the same number of neighbouring cells, that is, $|V_g|$ is independent of $g$. We are led to introduce a multiplication law for the elements of $\cal G$ and define a subset $S$ such that $g'=gh$ for every neighbour $g'$ of $g$. It follows that $E=\{(g,gh) | g\in {\cal G}, h \in S\}$, explicitly treating all neighbouring sets on equal footing. We also assume $\cal G$ to be closed under multiplication so that $\cal G$ forms a group finitely generated by $S$. The subset $S$ is often called the generating set of the group $\cal G$ in the framework of Cayley graphs \cite{white2001}. By construction the inverse element $h^{-1}$ of $h$ must necessarily be in $S$ as the consideration of a primitive cell centred on $g'$ immediately reveals. In the following it will be useful to write $S=S_+ \cup S_-$, where $S_-$ denotes the set containing the inverse elements of $S_+$. Note that this decomposition is not unique. In this paper we only consider the case of a body-centred cubic system, where the set $S_+$ contains the vertices of a regular tetrahedron and $S_-$ contains the vertices of the corresponding dual tetrahedron (see Fig.~\ref{fig:tetra}).

Third, we assume the cellular automaton to be linear. Physically, it means that no interaction between fermionic fields is allowed and we can focus on the dynamics of a single particle. In fact, the term Quantum Cellular Automata is often reserved for the dynamics of interacting fields while Quantum (Random) Walks (QW) is used for the dynamics of a single particle \cite{aharonov1993,kempe2003,werner2012}. As our general framework started with an abstract QCA, we will keep the term QCA throughout this article, although we ultimately examine a single particle. As often with QW \cite{lopez2006}, we can define the single-particle Hilbert space ${\cal H}={\cal H}_{\cal G} \otimes {\cal H}_F$ where the space ${\cal H}_{\cal G}$ is used to describe the position of the particle while the space ${\cal H}_F$ is used to describe its internal degree of freedom. Since we are interested in two-dimensional automata, we simply choose ${\cal H}_F=\mathbb{C}^2$. Linearity means that the vector $\psi_g$ is the linear combination of the linearly transformed vectors of its neighbouring vertices and itself, at the preceding time-step. Therefore, the discrete-time evolution of the vector $\psi_g(t)$ at vertex $g$ from step $t$ to step $t+1$ with transition matrices $A_{g,g'}$ can be written as
\begin{eqnarray}\label{first}
\psi_g(t+1)=\sum_{g' \in V_g} A_{g,g'} \psi_{g'}(t) + A_{g,g} \psi_g(t),
\end{eqnarray}
where $A_{g,g}$ represents the transition matrix at the centre of the cell. Using the definition of the generating set $S$, this evolution becomes
\begin{eqnarray}
\psi_g(t+1)=\sum_{h \in S} A_{g,gh}\psi_{gh}(t) + A_{g,g} \psi_{g}(t).
\end{eqnarray}

Finally, the above local evolution can be rewritten using the Hilbert space structure ${\cal H}={\cal H}_{\cal G} \otimes \mathbb{C}^2$. Now, we simply denote the transition matrices $A_{g,gh}$ as $A_h$, the neutral element of $\cal G$ as $e$, and we introduce the shift operator $T_h$ in the right regular representation that shifts a vertex $|x\rangle$ to the vertex ${T_{y}|x\rangle=|x y^{-1}\rangle}$. It follows that the neighbouring vertex $g'=gh$ is shifted by $T_h$ to the central vertex $g$. We finally obtain the evolution operator $A$ such that ${\psi(t+1)=A \psi(t)}$,  $\psi(t)$ in $\cal{H}$, as
\begin{eqnarray} \label{equation}
A=\sum_{h \in S} T_h \otimes A_h + \openone \otimes A_e.
\end{eqnarray}
Such an operator is often called the walk operator while the operator $A_e$ is often taken to be zero. In this paper, we do not assume $A_e$ to vanish.

Next, we demand the automaton to be isotropic, that is, all the directions on the lattice are equivalent. The corresponding mathematical definition is not trivial. We require the existence of a finite subgroup $L$ of the automorphism group $\textrm{Aut}(\cal G)$ of $\cal G$, transitive over $S_+$, and a unitary representation $U$ of $L$ such that, for all $l$ in $L$,
\begin{eqnarray}\label{iso}
A=\sum_{h \in \tilde{S}} T_h \otimes A_h=\sum_{h \in \tilde{S}} T_{l(h)} \otimes U_l A_h U_l^\dagger,
\end{eqnarray}
where we include both $S$ and the neutral element $e$ in the definition of the set $\tilde{S}$ and where $l(h)$ denotes the new vertex after action of $l$ on $h$. Note that the above definition does not require the isotropy group $L$ to include all the symmetries of the lattice. To ensure that all directions really are equivalent, an alternative definition of isotropy would be to require the isotropy group to be the full symmetry group of the lattice. Importantly, the above covariance of the automaton $A$ implies the invariance under conjugation of the two set of transition matrices $A_{h}$ and $A_{h^{-1}}$, that is,
\begin{eqnarray}\label{iso2}
A_{l(h^{\pm 1})}=U_l A_{h^{\pm 1}} U_l^\dagger,
\end{eqnarray}
for all $h$ in $S_+$ and $h^{-1}$ in $S_-$. Indeed, the choice of a regular representation implies that the $T_h$'s are linearly independent in the sense that if the action of a finite linear combination of the $T_h$'s on any site $|g\rangle$ of $\cal H_{\cal G}$ vanishes, then each coefficient in the linear combination must vanish. Consequently, the equality in Eq.~(\ref{iso}) must hold term by term in the finite sum over $\tilde{S}$. Due to the transitivity of $L$ over $S_+$, the two subsets $S_\pm$ remain unchanged under the action of the isotropic group $L$ and we end up with Eq.(\ref{iso2}). It also follows that the transition matrix $A_0$ at the centre of the cell must be invariant under conjugation. If the unitary representation $U$ of $L$ is irreducible, then Schur's Lemma ensures that $A_0$ is proportional to the identity matrix or zero. Let us emphasise here that isotropy as defined in Eq.~(\ref{iso}) is a very strong constraint indeed. Although the isotropy group $L$ will not be {\it a priori} specified,  covariance will be instrumental in characterising the QCA. Once the solutions are known, it will be possible to find their symmetries and therefore to identify the isotropy group $L$ and its unitary representation $U$. 

So far, the construction holds for any countable group $\cal G$ finitely generated by a subset $S$. We now fix $\cal G$ to be the abelian group of translations in the three-dimensional body-centred cubic lattice. This choice of group represents our discrete three-dimensional lattice. Mathematically, the choice of an abelian group is convenient as it allows to work in the Fourier space, also called the reciprocal lattice in crystallography. The irreducible representations of an abelian group are one-dimensional and unitary. Thus, we can now label the group elements of $\cal G$ as integer vectors in ${\mathbb Z}^3$, denoted in bold letters, and use the sum notation for the commutative group law so that
\begin{eqnarray}
T_{\bold y} | {\bold x}\rangle=| \bold{x-y}\rangle.
\end{eqnarray}
The periodicity of the lattice allows us to define the (continuous) Fourier-transformed basis $\{|\bold{k}\rangle\}$ of the (discrete) basis $\{|\bold{x}\rangle\}$ as
\begin{eqnarray}
| {\bold k}\rangle= \frac{1}{\sqrt{|{\cal B}|}} \sum_{\bold{x} \in \mathbb{Z}^3 } \textrm{e}^{i \bold{k}.\bold{x}} | \bold{x}\rangle
\end{eqnarray}
where $|\cal B|$  denotes the volume of the first Brillouin zone  $\cal B$ of the body-centred cubic lattice. This first Brillouin zone takes the shape of a rhombic dodecahedron that comprises all wave vectors $\bold{k}$ such that in cartesian coordinates $-\pi \leq \pm k_x \pm k_y \leq \pi$, $-\pi \leq \pm k_x \pm k_z \leq \pi$, and $-\pi \leq \pm k_y \pm k_z \leq \pi$ \cite{brillouin2001}. Importantly, the shift operators $T_{\bold h}$ are diagonal in the Fourier basis, that is, $T_\bold{h} |\bold{k}\rangle= e^{i \bold{h}.\bold{k}}|\bold{k}\rangle$ so that the unitary automaton takes the form
\begin{eqnarray}
A=\sum_{{\bold h} \in \tilde{S}} \int_{\cal B} d\bold{k} \, e^{i \bold{k}.\bold{h}} |\bold{k}\rangle \langle \bold{k}| \otimes A_{\bold h}.
\end{eqnarray}
The final form of a homogenous, local, isotropic, and unitary automaton on a body-centred cubic lattice then is
\begin{eqnarray}\label{Afinal}
A= \int_{\cal B} d\bold{k} \, |\bold{k}\rangle \langle \bold{k}| \otimes {\cal A}(\bold k)
\end{eqnarray}
where the operator
\begin{eqnarray}\label{Ak}
{\cal A}(\bold k)=\sum_{\bold{h} \in \tilde{S}} e^{i \bold{k}.\bold{h}} A_\bold{h}
\end{eqnarray}
is unitary for all wave vectors in the first Brillouin zone and the set of transition matrices $A_{\bold{h}}$, $\bold h$ in $S_+$, the set of transition matrices $A_{\bold{h}^{-1}}$, and $A_0$ are invariant under conjugation for some group of unitaries.

Interestingly, the unitary property of the automata $A$ further implies that the transition matrices must satisfy the conditions
\begin{eqnarray} \label{unitary1}
\sum_{\bold{h} \in \tilde{S}} A_\bold{h}^\dagger A_\bold{h}=\openone \quad \textrm{and} \,\,
\sum_{\bold{h''}=\bold{h}-\bold{h'}} A_\bold{h'}^\dagger A_\bold{h}=0.
\end{eqnarray}
Indeed, since the operators $T_{\bold h}$ are unitary, Eq.~(\ref{equation}) implies that
\begin{eqnarray}
A^\dagger A=\openone \otimes \sum_{\bold{h}} A_\bold{h}^\dagger A_\bold{h} + \sum_{\bold{h''=\bold{h}-\bold{h'}}} T_\bold{h''} \otimes A_\bold{h'}^\dagger A_\bold{h},
\end{eqnarray}
where each term in the second sum has to vanish due to the linear independence of the operators $T_{\bold{h}}$. Similarly, the condition $A A^\dagger=\openone$ immediately reads
\begin{eqnarray} \label{unitary2}
\sum_{\bold{h}  \in \tilde{S}} A_\bold{h} A_\bold{h}^\dagger=\openone \quad \textrm{and} \,\,
\sum_{\bold{h''}=\bold{h}-\bold{h'}} A_{\bold{h'}} A_\bold{h}^\dagger=0.
\end{eqnarray}

Homogenous, linear, local, isotropic, and unitary automata have another interesting property. Indeed, for any homogenous, linear, and local automaton, we have
\begin{eqnarray}
(\openone \otimes {\cal A}^\dagger(\bold k=\bold0)) \, A = \sum_{{\bold h} \in \tilde{S}} T_{\bold h} \otimes A'_{\bold h},
\end{eqnarray}
where $\sum_{{\bold h} \in \tilde{S}}  A'_{\bold h} =\openone$. In dimension $d$, the operator ${\cal A}(\bold k=\bold 0)$ is a dxd unitary transformation that is specified by $d^2$ real parameters. Therefore, the above equation defines a $d^2$-parameter family of automata constructed from an automaton $A$ such that $\sum_{{\bold h} \in \tilde{S}}  A_{\bold h} =\openone$ and multiplied by a transformation of the form $\openone \otimes V$. If, furthermore, the automaton is isotropic, then the operator ${\openone \otimes {\cal A}^\dagger(\bold k=0)}$ is invariant under the unitary conjugation of the isotropy group $L$. It follows that if the representation of $L$ is irreducible, then, by Schur's lemma, ${\cal A}^\dagger(\bold k=0)$ must be zero or a multiple of the identity. If the automaton is also unitary, it cannot vanish and the multiplicative constant must be a phase such that ${\cal A}(\bold k=0)$ reduces to the identity operator up to a physically-irrelevant phase. Therefore, if the isotropy group $L$ has an irreducible representation in two dimensions, the four-parameter family of QCA reduces to a single QCA. Here we see again how strong the isotropy constraint is. In the following, we do assume an automaton fulfilling the property ${\cal A}(\bold k=0)=\openone$. In other words, we impose the constraint
\begin{eqnarray}\label{id}
\textrm{C}_{\textrm{0}}: \quad \quad \, A_0+ \sum_\bold{h} A_\bold{h}+A_{-\bold{h}} =\openone.
\end{eqnarray}
Physically, this assumption is rather intuitive: If all directions are equivalent, it seems natural to require that the vertex remains unchanged when the wave vector of the particle is zero.
This represents our first constraint. In the following, we will  introduce three additional constraints coming from the unitary constraint. These four constraints will be enough to determine fully the searched automata.

\section{Derivation of the Weyl automata}\label{sectionWeyl}

We now start with the novelty of this article: A short and simple derivation of the Weyl automata. The main difference between D'Ariano and Perinotti's derivation in \cite{dariano2014} and that of the present paper is the use of Gram matrix. While D'Ariano and Perinotti express the transition matrices in various forms to tackle different constraints, we only consider their Pauli decomposition to make use of the notion of Gram matrix. It is this notion that allows a shorter and perhaps more transparent derivation. Another distinction concerns the transition matrix at the center of the body-centred cubic cell that we never assume to vanish.

We call a homogenous, local, isotropic, and unitary automaton, a Weyl automaton as we already know that such an automaton necessarily reduces to the Weyl equation in the limit of small wave vectors and continuous time \cite{birula1994}. To find such automata, we will examine the unitarity of the operator ${\cal A}(\bold k)$ defined in Eq.~(\ref{Ak}). But first, let us introduce two relevant Gram matrices. 

\subsection{Gram matrix}

The operator ${\cal A}(\bold k)$ can be written as
\begin{eqnarray}\label{automaton}
{\cal A}(\bold k)=A_0+ \sum_j e^{i k_j} A_j+e^{-i k_j} A_{-j},
\end{eqnarray}
with $k_j=\bold{k.h}_j$ and where the eight vectors $\bold{h}_{j}$ and $\bold{h}_{-j}$, $j=1,2,3,4$, correspond to the vertices of the regular tetrahedron $S_+$ and its dual $S_-$. Note that we now enumerate the vectors $\bold{h}$ as $\bold{h}_j$ and transition matrices $A_{\bold h}$ as $A_j$ to simplify the notations. The non-normalised vectors of these two tetrahedra are defined in some orthonormal basis as
\begin{eqnarray}\label{tetra}
\bold h_1=\begin{bmatrix} 1\\ 1\\1 \end{bmatrix}\!\!,
\,\,\bold h_2=\begin{bmatrix} 1\\ -1\\-1 \end{bmatrix}\!\!,
\,\, \bold h_3=\begin{bmatrix} -1\\ 1\\-1 \end{bmatrix}\!\!,
\,\,\bold h_4=\begin{bmatrix} -1\\ -1\\1 \end{bmatrix}\!\!, \nonumber
\end{eqnarray}
and
\begin{equation}
\bold{h}_{-j}=-\bold{h}_{j}.
\end{equation}
Our task is to find the nine 2x2 transition matrices $A_{\pm j}$ and $A_0$ in a simple and insightful manner.

We recall here that the Gram matrix of a collection of $n$ real vectors $\bold{v}$ is defined by the nxn symmetric matrix $G$ such that ${G_{jk}=\bold{v}_j.\bold{v}_k}$, with the usual definition of the dot product. In our case, the relevant vectors are complex, so we will define two Gram matrices denoted by $G^R$ and $G^C$ as follows
\begin{equation}\label{gram}
G^R_{jk}=\bold{v}_j.\bold{v}_k \quad \textrm{and} \quad G^C_{jk}=\bold{v}^*_j.\bold{v}_k.
\end{equation}
Note that only $G^C$ is positive semi-definite as the simple dot product in the definition of $G^R$ does not correspond to an inner product. Furthermore, for a set of four vectors, we can define a 3x4 matrix $B$ whose columns are the vectors $\bold{v}_i$s so that the two Gram matrices defined above take the simple form $G^R=B^tB$ and $G^C=B^\dagger B$, where $t$ denotes transposition while $\dagger$ denotes Hermitian conjugation.

The regular tetrahedron defined in Eq.~(\ref{tetra}) can be fully characterised in a coordinate-free manner by the relation
\begin{equation}
\bold{h}_j.\bold{h}_k=4 \delta_{jk}-1,
\end{equation}
where $\delta_{jk}$ is the Kronecker delta. Interestingly, these relations can be grouped in the Gram matrix
\begin{equation}\label{G}
G=T^t T=\begin{bmatrix} 3&-1&-1&-1\\ -1&3&-1&-1\\-1&-1&3&-1\\-1&-1&-1&3 \end{bmatrix}
\end{equation}
where \begin{equation}\label{T}
T=\begin{bmatrix} 1&1&-1&-1\\ 1&-1&1&-1\\1&-1&-1&1\end{bmatrix}
\end{equation}
is the matrix containing the four vectors of the non-normalised tetrahedron. Finally, the four vectors of the regular tetrahedron in three-dimensional space verifies the simple linear-dependence relation $\sum_{j=1}^4 \bold{h}_j=\bold{0}$. In other words, the sum of all the terms in any row of the Gram matrix $G$ vanishes.

\subsection{Unitary constraint}
The  unitary constraint ${\cal A}(\bold k){\cal A}^\dagger(\bold k)=\openone$ is a sum of eighty one terms of the form $e^{\pm i k_i} e^{\mp i k_j} A_{\pm i} A^\dagger_{\pm j}$, $e^{\mp i k_j} A_{0} A^\dagger_{\pm j}$, $e^{\pm i k_j}  A_{\pm j}A_{0}^\dagger$, and $A_0A_{0}^\dagger$. These terms can be grouped depending on their relative phase as given in Eq.~(\ref{unitary2}) and each group must vanish. Among these groups, three will suffice to characterise the automaton. More specifically,  we choose the three groups with relative phases $e^{2 i k_j}$, $e^{i (k_i-k_j)}$, and $e^{-i k_j}$, respectively. Therefore, we end up with the three constraints
\begin{eqnarray}
\textrm{C}_{\textrm{1a}}: &\quad& A_j A^\dagger_{-j}=0, \label{C1a}\\
\textrm{C}_{\textrm{2a}}: &\quad& A_i A^\dagger_j +A_{-j} A^\dagger_{-i}=0, \, i \neq j,\label{C2a}\\
\textrm{C}_{\textrm{3a}}: &\quad& A_0 A^\dagger_j +A_{-j} A^\dagger_{0}=0,\label{C3a}
\end{eqnarray}
and similarly for  ${\cal A}^{\dagger}(\bold k){\cal A}(\bold k)=\openone$,
\begin{eqnarray}
\textrm{C}_{\textrm{1b}}: &\quad& A^\dagger_j A_{-j}=0,\label{C1b}\\
\textrm{C}_{\textrm{2b}}: &\quad& A^\dagger_i A_j +A^\dagger_{-j} A_{-i}=0, \, i \neq j, \label{C2b}\\
\textrm{C}_{\textrm{3b}}: &\quad& A^\dagger_0 A_j +A^\dagger_{-j} A_{0}=0. \label{C3b}
\end{eqnarray}

Together with the constraint $\textrm{C}_{\textrm{0}}$ of Eq.~(\ref{id}) repeated here for convenience
\begin{eqnarray} \label{c0}
\textrm{C}_{\textrm{0}}: \quad \quad \, A_0+ \sum_j A_j+A_{-j}=\openone,
\end{eqnarray}
we now have a total of four constraints that we can start exploiting to find the Weyl automata.
 
\subsection{From two tetrahedra to one tetrahedron}\label{subsection3b}

We express the 2x2 complex matrices $A_0$ and $A_{\pm j}$ in terms of the three Pauli matrices $\sigma_1$, $\sigma_2$, and $\sigma_3$ defined as
\begin{equation}
\sigma_1=\begin{bmatrix} 0&1\\ 1&0 \end{bmatrix}\,,
\quad
\sigma_2=\begin{bmatrix} 0&-i\\ i&0 \end{bmatrix}\,,
\quad
\sigma_3\begin{bmatrix} 1&0\\0&1 \end{bmatrix},
\end{equation}
together with the identity matrix $\openone$. We will see that in the case of an isotropic automaton the transition matrices $A_{\pm j}$ can be written as
\begin{equation}\label{form}
A_j=\frac{\alpha}{2}(\openone + \bold{a}_j . \sigma) \quad \textrm{and} \quad
A_{-j}=\frac{\beta}{2}(\openone - \bold{a}^*_{j} . \sigma),
\end{equation}
where
\begin{equation}\label{a1}
\bold a_j.\bold a_j=1,
\end{equation}
and $\alpha$ and $\beta$ are two complex numbers different from zero. In other words, the transition matrices $A_{\pm j}$ have non vanishing traces and  are such that ${\bold a}_{-j}=-{\bold a}_j^*$. The immediate consequence is that we only need to focus on the four complex vectors $\bold{a}_j$ as well as the two complex factors $\alpha$ and $\beta$ to solve our problem. Let us now prove these claims.

The two 2x2 complex matrix $A_{ j}$ and $A_{- j}^\dagger$ can always be written as
\begin{equation}\label{decomposition}
A_{j}=a_{j} \openone + {\bold a}_{j} . \sigma
\quad \textrm{and} \quad
A_{-j}^\dagger=a^*_{- j} \openone + \bold{a}_{- j}^*\!\!. .\sigma
\end{equation}
with $a_{\pm j} \in \mathbb C$, ${\bold a}_{\pm j} \in \mathbb C^3$. The two constraints $\textrm{C}_{1\textrm{a}}$ and $\textrm{C}_{1\textrm{b}}$ of Eqs.~(\ref{C1a}) and (\ref{C1b}) yield the three conditions
\begin{eqnarray}
a_j a_{- j}^*+\bold a_j.\bold a_{-j}^*=0,\label{MN1}\\
a_{-j}^* \bold a_j+a_j \bold{a}_{- j}^* \pm i \, \bold{a}_{j} \,\textrm{x} \, \bold{a}_{-j}^*=0.\label{MN2}
\end{eqnarray}
There are three possible cases for the above system: First, none of the two matrices $A_{ j}$ and $A_{- j}$ has a trace equal to zero. Second, only one matrix has a trace equal to zero. Third, both matrices have a trace equal to zero. In the first case, upon rescaling the complex vectors $\bold a_j$ and $\bold a_{-j}$ and using a conventional factor $1/2$, the two matrices $A_j$ and $A_{-j}$ take the simple form given in Eqs.~({\ref{form}}) and (\ref{a1}). Of course, the coefficient $\alpha$ ($\beta$) must be the same for all transition matrices $A_j$ ($A_{-j}$) since they are unitarily conjugated under the action of the isotropy group and, therefore, their traces must be equal. Given the four complex vectors $\bold{a}_j$, we define the two Gram matrices $G^R$ and $G^C$ as $G^R_{jk}=\bold{a}_j.\bold{a}_k$ and $G^C_{jk}=\bold{a}^*_j.\bold{a}_k$, respectively. The full characterisation of these two Gram matrices will specify the four vectors $\bold{a}_j$, up to rotations. Such a freedom on the vector $\bold{a}_j$ corresponds to the definition of an automaton up to unitary conjugation. Interestingly, Eq.~(\ref{a1}) already fixes the diagonal of the Gram matrix $G^R$: All the diagonal entries of the Gram matrix $G^R$ must be equal to $1$. Similarly, since $A_1$ and $A_2$ are unitarily conjugated,  $A_1^\dagger A_1$ and $A_2^\dagger A_2$ must be unitarily conjugated too so that their traces must be equal, i.e.\ $\bold{a}^*_1.\bold{a}_1=\bold{a}^*_2.\bold{a}_2$. Therefore, our isotropic automaton is such that for all $j$, $\bold{a}^*_j.\bold{a}_j=c$ where $c$ is a positive number: All the diagonal entries of the Gram matrix $G^C$ must be equal to $c$.

In the second case, it immediately follows that the matrix with vanishing trace must completely vanish. Let us say that this vanishing matrix is $A_{-j}$. Since all the transition matrices  $A_{-j}$ are unitarily conjugated, all of them must vanish. Thus, the constrains $\textrm{C}_{2\textrm{a}}$ and $\textrm{C}_{2\textrm{b}}$ of Eqs.~(\ref{C2a}) and (\ref{C2b}) reduce to ${A_iA_j^\dagger=A_i^\dagger A_j=0}$ and we must have $\bold a_j=-\bold a_i^*$ together with $\bold a_i.\bold a_i=1$. Note that there is no other case as, by assumption, the matrices $A_i$ have a non-zero trace. Since the constrains $\textrm{C}_{2\textrm{a}}$ and $\textrm{C}_{2\textrm{b}}$ must be fulfilled for any pair $i \neq j$, all four vectors $\bold{a}_j$ must necessarily be equal, imaginary, and such that $\bold{a}_i.\bold{a}_j^*=-1$ for all pairs $i,j$ (even for $i=j$). This contradicts the fact that the Gram matrix $G^C_{jk}=\bold{a}^*_j.\bold{a}_k$ must be positive semi-definite.

In the third case, all transition matrices have zero trace. It follows from Eqs.~(\ref{MN1}) and (\ref{MN2}) that ${\bold{a}_j.\bold{a}_{-j}^*=0}$ and ${\bold{a}_{j} \,\textrm{x} \, \bold{a}_{-j}^*=0}$. Thus, there exists non-zero complex numbers $\lambda_j$ such that ${\bold{a}_{-j}^*=\lambda_j \bold{a}_{j}}$ and $\bold{a}_j.\bold{a}_j=0$. The constrain $\textrm{C}_{2\textrm{a}}$ now reads ${(1+\lambda_i \lambda_j^*) \bold{a}_i.\bold{a}_j^*=0}$ and $(1-\lambda_i \lambda_j^*) \bold{a}_i {\textrm x}\, \bold{a}_j^*=0$ such that necessarily $\bold{a}_i.\bold{a}_j^*=0$, for all $i \neq j$. Thus, the corresponding Gram matrix $G^C$ contains only vanishing off-diagonal elements. Importantly, $G^C$ is the  Gram matrix of four vectors in three dimensions, therefore its determinant must vanish. Since all its diagonal elements must be identical because of the isotropy group, it follows that $G^C$ must entirely vanish so that all transition matrices $A_{\pm j}$ are zero. The constraint $\textrm{C}_0$ of Eq.(\ref{c0}) directly implies that $A_0$ and the automaton ${\cal A}(\bold k)$ itself must be the identity matrix. This completes the very special case where the traces of all transition matrices vanish.

In conclusion, all the transition matrices $A_{\pm j}$ have either zero or non-zero traces while a more asymmetrical case is forbidden. We now focus on the non-trivial case where the transition matrices have non-vanishing traces and can be written in the form given in Eqs.~(\ref{form}) and (\ref{a1}) with non-zero $\alpha$ and $\beta$.

\subsection{Characterisation of the Gram matrix $G^C$}
The two constraints $\textrm{C}_{2\textrm{a}}$ and $\textrm{C}_{2\textrm{b}}$ will fix all the off-diagonal entries of the Gram matrix $G^C$ and most of the off-diagonal terms of the Gram matrix $G^R$. Let us now proceed with solving the constraint $\textrm{C}_2$. The identity $A_i A^\dagger_j +A_{-j} A^\dagger_{-i}=0$, $i \neq j$ reads
\begin{eqnarray} 
(|\alpha|^2+|\beta|^2) (1+\bold{a}_i . \bold{a}^*_j)=0, \label{originalC2a} \\
(|\alpha|^2-|\beta|^2) (\bold{a}_i + \bold{a}^*_j + i \, \bold{a}_i \,\textrm{x} \, \bold{a}^*_j)=0. \label{originalC2a2}
\end{eqnarray}
Similarly, the identity $A^\dagger_i A_j +A^\dagger_{-j} A_{-i}=0$, $i \neq j$ reads
\begin{eqnarray}
(|\alpha|^2+|\beta|^2) (1+\bold{a}_i . \bold{a}^*_j)=0, \label{originalC2b}\\
(|\alpha|^2-|\beta|^2) (\bold{a}_i + \bold{a}^*_j - i \, \bold{a}_i \,\textrm{x} \, \bold{a}^*_j)=0.\label{originalC2b2}
\end{eqnarray}
Since $\alpha$ and $\beta$ are non zero, it immediately follows that
\begin{eqnarray} \label{GC_diago}
\bold{a}_i . \bold{a}^*_j=-1,
\end{eqnarray}
such that the Gram matrix $G^C$ is of the form
\begin{equation}\label{spec}
G^C=\begin{bmatrix} c&-1&-1&-1\\ -1&c&-1&-1\\-1&-1&c&-1\\-1&-1&-1&c \end{bmatrix}.
\end{equation}
Therefore, $G^C$ is real and symmetric and it can be diagonalised using the rotation matrix
\begin{equation}\label{rot}
R=\frac{1}{2}\begin{bmatrix} 1&1&1&1\\ 1&1&-1&-1\\ 1&-1&1&-1\\1&-1&-1&1\end{bmatrix}
\end{equation}
which is nothing but the matrix $T$ of Eq.~(\ref{T}) with an additional first row of ones and a normalising factor $1/2$. The eigenvalues of $G^C$ then are $\{1+c,1+c,1+c,-3+c\}$. Since $G^C$ is the Gram matrix of four vectors in three dimensions, its determinant must vanish. In other words, $c$ can only be -1 or 3. The number $c=\bold{a}_i .\bold{a}^*_i$ being positive, we end up with
\begin{equation}\label{spec}
G^C=\begin{bmatrix} 3&-1&-1&-1\\ -1&3&-1&-1\\-1&-1&3&-1\\-1&-1&-1&3 \end{bmatrix},
\end{equation}
which is nothing but the Gram matrix $G$ of the tetrahedron given in Eq.~(\ref{G}). In other words, we have $G^C=T^t T$. This identity suggests a link between the transition matrices $A_j$ and the regular tetrahedron. Before turning our attention to the off-diagonal elements of $G^R$, let us note that the two equations Eqs.~(\ref{originalC2a2}) and (\ref{originalC2b2}) imply 
\begin{equation}\label{mod}
|\alpha|=|\beta|.
\end{equation}
If we assume otherwise, the four vectors $\bold{a_j}$ would all be equal, so that $\bold{a}_i . \bold{a}^*_j=\bold{a}_i . \bold{a}^*_i \geq 0$. This would contradict the additional requirement that $\bold{a}_i . \bold{a}^*_j=-1$.

So far, we haven't considered the transition matrix at the centre of the primitive cell. We shall soon see that $A_0$ must vanish. Using the decomposition of Eq.~(\ref{form}) for the transition matrices $A_{\pm j}$ together with $A_0=a_0 \openone + \bold a . \sigma$, we can rewrite the constraint  $\textrm{C}_0$ in Eq.~(\ref{c0}) as
\begin{equation}
\left(a_0+2 (\alpha+\beta) \right)\openone + \left(\bold{a}+ \frac{\alpha}{2} \sum_j \bold{a}_j - \frac{\beta}{2} \sum_j \bold{a}^*_j\right)\!\!.\sigma=\openone.
\end{equation}
It immediately follows that
\begin{eqnarray}
a_0+2 (\alpha+\beta) &=&1,\label{ww}\\
\bold{a}_k . \bold{a}+\frac{\alpha}{2} \sum_j \bold{a}_k . \bold{a}_j&=&0 \label{qq}
\end{eqnarray}
since $\sum_j \bold{a}_k . \bold{a}^*_j=0$, that is, the sum of all the terms in any row of the Gram matrix $G^C$ vanishes. We now need the constraint $\textrm{C}_3$ that corresponds to the three identities
\begin{eqnarray} 
\alpha^* a_0+\beta a_0^*+(\alpha^* \bold{a}-\beta \bold{a}^*).\bold{a}_j^*&=&0,\label{aa}\\
\alpha^* \bold{a}+\beta \bold{a}^*+(\alpha^* a_0-\beta a_0^*)\bold{a}_j^*&=&0, \label{ab}\\
(\alpha^* \bold{a}+\beta \bold{a}^*) \,\textrm{x} \, \bold{a}_j&=&0 \label{ac}.
\end{eqnarray} 
In particular, Eq.~(\ref{ab}) implies
\begin{eqnarray} 
\alpha^* \bold{a}.\bold{a}^*_j+\beta \bold{a}^*.\bold{a}^*_j+(\alpha^* a_0-\beta a_0^*)=0
\end{eqnarray}
since $\bold{a}_j.\bold{a}_j=1$. It follows from this equation together with Eq.~(\ref{aa}) that
\begin{eqnarray} \label{A00}
a_0= \bold{a}.\bold{a}_j=\bold{a}^*.\bold{a}_j=0.
\end{eqnarray}
Now, we can insert these results in Eqs.~(\ref{ww}) and (\ref{qq}) to obtain
\begin{eqnarray}
\alpha+\beta &=&\frac{1}{2} \label{out}\\
\sum_j \bold{a}_k . \bold{a}_j&=&0 \label{lin}.
\end{eqnarray}
The later identity means that the sum of all the terms in any row of the Gram matrix $G^R$ vanishes. We will see in the next subsection that the four vectors $\bold{a}_j$ span a three-dimensional space and therefore $\bold{a}.\bold{a}_j=0$ in Eq.~(\ref{A00}) implies $\bold{a}=0$ so that $A_0$ must vanish.

\subsection{Characterisation of the Gram matrix $G^R$}

We now are ready to characterise completely $G^R$ and $A_0$ and to find the Weyl automata. To do so, we do not need to solve any further the constraints $\textrm{C}_2$ or $\textrm{C}_3$. Instead, we simply exploit Eq.~(\ref{lin}). These are four linear constraints for six unknowns since the diagonal terms of the Gram matrix $G^R$ are already fixed to 1 as given in Eq.~(\ref{a1}). Therefore we finds that $G^R$ only depends on two unknowns, say $x=\bold{a}_1.\bold{a}_2$ and $y=\bold{a}_1.\bold{a}_4$, and it takes the form
\begin{equation}\label{2unknowns}
G^R=\begin{bmatrix} 1&x&-\!1\!-\!x\!-\!y&y&\\ x&1&y&-\!1\!-\!x\!-\!y\\-\!1\!-\!x\!-\!y&y&1&x\\y&-\!1\!-\!x\!-\!y&x&1 \end{bmatrix}.
\end{equation}
Next, we observe that the Gram matrix $G^R$ can be diagonalised by the very same rotation $R$ as $G^C$. Its eigenvalues are $\{0,2(1+x),-2(x+y),2(1+y)\}$ such that it can be written as
\begin{equation}\label{matD}
G^R=T^t \begin{bmatrix} (1+x)/2&0&0\\0&-(x+y)/2&0\\0&0&(1+y)/2 \end{bmatrix} T.
\end{equation}
This identity leaves little freedom to the unkonwns $x$ and $y$. Indeed, by definition of the two Gram matrices $G^R$ and $G^C$, there exists a 3x4 matrix $B$ such that $G^R=B^tB$ and $G^C=B^\dagger B$. Since we already know that $G^C=T^t T$ with ${T^t=T^\dagger}$, there must exist a 3x3 unitary matrix $Z$ such that $B=ZT$ and we can write the Gram matrix $G^R$ as $G^R=T^t Z^t ZT$ or simply ${G^R=T^t W T}$ where $W$ is a 3x3 symmetric unitary matrix. If we denote the diagonal matrix in Eq.(\ref{matD}) as $D$, we now have ${T^t D T=T^t W T}$. We can multiply each side of this identity by $T/4$ on the left and $T^t/4$ on the right to obtain ${D=W}$. Since $D$ is diagonal while $W$ is unitary, these two matrices can only be equal if they are diagonal with only phases on the diagonal. In other words, we simply have ${|(1+x)/2|=|(x+y)/2|=|(1+y)/2|=1}$.

It follows that the pair $(x,y)$ must necessarily be equal to $(1,-3)$, $(1,1)$, or $(-3,1)$. The three corresponding Gram matrices $G^R$ are
\begin{eqnarray}
G^R_1=\begin{bmatrix} 1&1&1&-3\\ 1&1&-3&1\\1&-3&1&1\\-3&1&1&1 \end{bmatrix} \label{sol1},\\
G^R_2=\begin{bmatrix} 1&1&-3&1\\ 1&1&1&-3\\-3&1&1&1\\1&-3&1&1 \end{bmatrix} \label{sol2},\\
G^R_3=\begin{bmatrix} 1&-3&1&1\\ -3&1&1&1\\1&1&1&-3\\1&1&-3&1 \end{bmatrix} \label{sol3}.
\end{eqnarray}
For the sake of completeness, we write the Gram matrices $G^C$ and $G^R$ in terms of the matrix $T$ containing the four vectors $\bold{h}_j$ of the regular tetrahedron. They are
\begin{eqnarray}\label{solution}
G^C&=&T^T T, \nonumber\\
G_1^R&=&T^T\begin{bmatrix} 1&0&0\\ 0&1&0\\0&0&-1 \end{bmatrix}T, \nonumber\\
G_2^R&=&T^T\begin{bmatrix} 1&0&0\\ 0&-1&0\\0&0&1 \end{bmatrix}T,\\
G_3^R&=&T^T\begin{bmatrix} -1&0&0\\ 0&1&0\\0&0&1 \end{bmatrix}T.\nonumber
\end{eqnarray}
Therefore, there are six possible matrices $B$ whose columns are the searched vectors $\bold{a}_j$. They are
\begin{eqnarray}
B_{1 \pm}=\begin{bmatrix}  1&0&0\\0&1&0\\0&0&\pm i \end{bmatrix} T,\label{sol1}\\
B_{2 \pm}=\begin{bmatrix} 1&0&0\\0&\pm i&0\\0&0&1 \end{bmatrix} T,\label{sol2}\\
B_{3 \pm}=\begin{bmatrix} \pm i&0&0\\0&1&0\\0&0&1 \end{bmatrix} T\label{sol3}.
\end{eqnarray}
We notice that $B_{j+}=B_{j-}^*$, $j=1,2,3$. Importantly, we now know that the four vectors $\bold{a}_j$ span a three dimensional space. Therefore, we can come back to Eq.~(\ref{A00}) and conclude that $\bold{a}.\bold{a}_j=0$ is only possible when $\bold{a}=0$ so that $A_0$ must vanish.

Importantly, the six automata corresponding to these six sets of vectors $\bold{a}_j$ are not all unitarily-inequivalent. To see that, let us find their spectra. Since the modulus of the determinant of a unitary matrix must be equal to 1 and we already know that $|\alpha|=|\beta|$, a direct calculation of the determinant of $\cal A(\bold k)$ shows that
\begin{eqnarray}
|\alpha|^2=|\beta|^2=\frac{1}{8}.
\end{eqnarray}
As we also know from Eq.~(\ref{out}) that $\alpha+\beta=1/2$, we necessarily have
\begin{equation}\label{alpha}
\alpha=\frac{1\pm i}{4}=\beta^*,
\end{equation}
leading to twelve possible automata. The spectra of these automata immediately follows from their traces. It turns out that there are only two different spectra $\{\textrm{e}^{i w_{\pm}}, \textrm{e}^{-i w_{\pm}}\}$, where
\begin{equation}\label{alpha}
\cos w_{\pm}=(\cos k_x \cos k_y \cos k_z \pm \sin k_x \sin k_y \sin k_z),
\end{equation}
$\bold k=\{k_x,k_y,k_z\}$. The $\pm$ sign in the above formula only depends on the choice $\alpha=(1\pm i)/4$. Two unitarily-inequivalent automata can conveniently be chosen to be $A(\bold{k})$ and $A^\dagger(\bold{-k})$, where the first automaton $A(\bold{k})$ is defined with ${\alpha=(1+i)/4}$ and the four vectors $\bold{a}_j$ of any $B_{i\pm}$ in Eqs.~(\ref{sol1})-(\ref{sol3}).

This completes the derivation of the homogenous, local, isotropic, and unitary two-dimensional automaton. We have found two unitarily-inequivalent automata
\begin{eqnarray} \label{W1}
{\cal A} (\bold{k})&=& \sum_j e^{i k_j} A_j +e^{-i k_j} A_{-j}
\end{eqnarray}
and $A^\dagger(\bold{-k})$, constructed from the four vectors $\bold{h}_j$ of the regular tetrahedron with a multiplicative phase $i$ in front of one of their three components. Since the transition matrices are of the form $A_j \propto (\openone + \bold{a}_j.\sigma)$ where the vectors $\bold{a}_j$ essentially are the vectors $\bold{h}_j$, we say that the transition matrices $A_{\pm j}$ essentially are the matrix representation of the vertices of the lattice's primitive cell. Finally, let us remember that we assumed the existence of an irreducible representation of the isotropy group in two dimensions. This isotropy group $L$ turns out to be the group of rotations of angle $\pi$ around the three coordinate axes, a rather simple group whose unitary irreducible representation $U$ in two dimensions is $\{\openone, \sigma_1, \sigma_2, \sigma_3\}$. 

\subsection{Continuum limit}
As proven in \cite{birula1994}, the found automata should reduce to the Weyl equation in the limit of small wave vectors $\bold{k}$ and continuous time. We can quickly verify it here. Indeed, we have for $|\bold{k}|<<1$,
\begin{eqnarray}
{\cal A} (\bold{k})&=& \alpha \sum_j (1+i k_j) \frac{1}{2} (\openone + \bold{a}_j.\sigma) \\
&+& \alpha^* \sum_j (1-i k_j) \frac{1}{2} (\openone - \bold{a}^*_j.\sigma).\nonumber
\end{eqnarray}
We can choose the automata  $A(\bold{k})$ to be given by ${\alpha=(1+i)/4}$ and the four vectors $\bold{a}_j$ of any $B_{j-}$ in Eqs.~(\ref{sol1})-(\ref{sol3}). Then, a direct calculation shows that the two unitary-inequivalent automata $A(\bold{k})$ and $A^\dagger(\bold{-k})$ reduce to
\begin{eqnarray} \label{W2}
{\cal A} (\bold{k})=\openone + i\, \bold{k}.\sigma.
\end{eqnarray}
If, instead, one chooses ${\alpha=(1+i)/4}$ together with the four vectors $\bold{a}_j$ of any $B_{j+}$, we would obtain the above equation for the wave vector $-\bold{k}$ and rotated by an element of the isotropy group. Finally, to obtain the continuum limit of our automata, we define the Hermitian operator $H(\bold k)$ of the unitary transformation ${\cal A} (\bold{k})=\textrm{e}^{-iH(\bold{k})}$ for the discrete time evolution ${|\psi(t+1)\rangle={\cal A} (\bold{k}) |\psi(t)\rangle}$ so that for a small-amplitude Hamiltonian ${A=\openone - i H(\bold{k})}$, $H(\bold k)=-\bold{k}.\sigma$. We then identify $H(\bold{k})$ with the Hamiltonian ${\cal H}$ of the unitary transformation ${U(t)=\textrm{e}^{-i {\cal H} t}}$ for the continuous time evolution ${i\partial_t \psi(t)={\cal H} \psi(t)}$. It follows that ${{\cal H}=-\bold{k}.\sigma}$. Therefore, in the limit of small wave vector and continuous time, the two homogenous, local, isotropic, and unitary two-dimensional automata $A(\bold{k})$ and $A^\dagger(\bold{-k})$ reduce to the Weyl equation in momentum space
\begin{eqnarray}\label{W3}
i\partial_t \psi=-\bold{k}.\sigma \psi.
\end{eqnarray}
Let us add for the sake of completeness that there are two Weyl equations discussed in the literature and often called right- and left-handed Weyl equations. They would correspond to the two Weyl automata $A(\bold{k})$ and $A^\dagger(\bold{k})$, not $A^\dagger(\bold{-k})$. In fact, in the four-dimensional case below we will find a QCA that corresponds to the coupling of two Weyl automata, one right-handed and one left-handed, as one would expect from the Dirac equation. Note also that rigorous continuum limits for quantum walks were discussed in \cite{arrighi2014}.

\section{Derivation of the Dirac automata}

With the two-dimensional case completed, we briefly turn our attention to the four-dimensional case. Four-dimensional homogenous, local, isotropic, and unitary automata on a three-dimensional lattice have been investigated in \cite{birula1994} and \cite{dariano2014}. Starting from a special form of four-dimensional automata, two families were found, both of them reducing to the Dirac equation in a suitable limit. In this paper, we use a different starting point but end up with the same two families of, so called, Dirac automata. More specifically, D'Ariano and Perinotti in \cite{dariano2014} started from two Weyl automata that they coupled. The coupling was then determined by the unitary constraint. In the present paper, we will try to solve the unitary constraint by restricting the possible solutions to a simple form inspired by the two-dimensional solutions but without assuming two Weyl automata. After some algebra, we will find two coupled Weyl automata.

Given the large dimension of the space where the solutions are to be found (i.e.\ sixteen real dimensions), we only consider a special class of automata. In the two-dimensional case, we discovered that the transition matrices $A_{\pm j}$ can essentially be thought as the matrix representation of the vertices of the lattice's primitive cell. Therefore, we keep the same structure but now use a four-dimensional representation of the Pauli algebra to find solution automata in the four-dimensional case. Such a representation can be achieved with the help of the four four-dimensional Gamma matrices: We simply replace the two-dimensional Pauli matrices $\sigma_i$ by the product $\gamma_i \gamma_0$, where the four Gamma matrices in the Weyl representation are defined as
\begin{equation}
\gamma_0=\begin{bmatrix} 0&\openone\\ \openone&0 \end{bmatrix}\,,
\quad
\gamma_i=\begin{bmatrix} 0&\sigma_i\\ -\sigma_i&0 \end{bmatrix}.
\end{equation}
Importantly, we also have $\openone=\gamma_0 \gamma_0$ such that our two-dimensional automata $\cal A(\bold k)$ given in Eq.~(\ref{automaton}) can be directly written in four dimensions. In the following, we denote all relevant four-dimensional matrices with a $B$ instead of an $A$. In other words, we now consider the matrices ${\cal B}(\bold{k})$, $B_{\pm j}$, and $B_0$ instead of ${\cal A}(\bold{k})$, $A_{\pm j}$, and $A_0$. We write
\begin{eqnarray}
{\cal B}(\bold{k})&=& B_0+\alpha \sum_j e^{i k_j} \frac{1}{2} (\openone+ \bold{a}_j.\gamma\gamma_0) \\
&+& \beta \sum_j e^{-i k_j} \frac{1}{2} (\openone - \bold{a}^*_j.\gamma\gamma_0).\nonumber
\end{eqnarray}

We choose to keep the four vectors $\bold{a}^*_{-j}=-\bold{a}_j$ and $|\alpha|=|\beta|$ from the two-dimensional results. However, $\alpha$, $\beta$, and $B_0$ are no longer known. These choices define a special class of automata, so that there may exist more general solutions than those we are about to find.

It is convenient to multiply the transition matrices $B_0$ and $B_j$ on the left by the full-rank matrix $\gamma_0$. This will simplify our calculations. We now have
\begin{eqnarray}
{\cal B}'(\bold{k})&=& B'_0+ \sum_j e^{i k_j} B'_j +  \sum_j e^{-i k_j} B'_{-j}.\nonumber
\end{eqnarray}
with $B'_j= \alpha/2(\gamma_0- \bold{a}_j.\gamma)$ and $B'_{-j}=\beta/2 (\gamma_0 + \bold{a}^*_j.\gamma)$. Interestingly, we see that $\alpha^* B'_{-j}=\beta B_j^{'\dagger}$ since ${\bold{a}_{-j}=-\bold{a}^*_j}$ while $\gamma_i^\dagger=-\gamma_i$ and $\gamma_0^\dagger=\gamma_0$. Second, the transition matrix $B'_0$ must satisfy the constraint $\textrm{C}_3$, so that we have
\begin{eqnarray}
\beta^*B'_0 B'_j &+& \alpha B'_{j}  B_0^{'\dagger}=0,\\
\alpha B_0^{'\dagger} B'_j &+& \beta^* B'_{j} B'_0=0.
\end{eqnarray}

An obvious solution to this system is $\beta^*=\alpha$ together with $B'_0=i r \openone$ where $r$ is a real number. To find $\alpha$ and $\beta$, we use the unitary constraint ${\cal B}'(\bold{k}){\cal B}^{'\dagger}(\bold k)=\openone$ that must be valid for all wave vectors $\bold{k}$. When the wave vector $\bold{k}$ vanishes, we find $r^2+16\textrm{Re}( \alpha)^2=1$. When we choose $\bold{k}=(\pi/2,\pi/2,\pi/2)^t$, we obtain ${r^2+16\textrm{Im}( \alpha)^2=1}$. It immediately follows that $\alpha=s(1\pm i)/4$, where $s$ is a real number such that $r^2+s^2=1$. Therefore, we obtain the automata
\begin{eqnarray}\label{bob}
{\cal B}'(\bold{k},s,\pm)&=& \pm i \sqrt{1-s^2} \openone \nonumber \\
&+& s \left( \sum_j e^{i k_j} B'_j + \sum_j e^{-i k_j} B'_{-j} \right),
\end{eqnarray}
where $B'_j=\alpha/2 (\gamma_0- \bold{a}_j.\gamma)$, $B'_{-j}=\alpha^*/2 (\gamma_0 + \bold{a}^*_j.\gamma)$, ${\alpha=(1\pm i)/4}$ and $s$ is a real parameter. Note that we now use $\alpha$ to denote $(1\pm i)/4$, instead of $s(1\pm i)/4$ as $s$ appears explicitly in Eq.~(\ref{bob}). Upon re-multiplying ${\cal B'}(\bold{k},s,\pm)$ by $\gamma_0$ on the left, these automata finally take the form
\begin{eqnarray}\label{sol4dim}
{\cal B}(\bold{k},s,\pm)=\begin{bmatrix} s A(\bold{k})&  \pm i \sqrt{1-s^2} \openone\\   \pm i \sqrt{1-s^2} \openone& s A^\dagger(\bold k)\end{bmatrix}.
\end{eqnarray}
In fact, one can go from an automata with parameter $+\sqrt{1-s^2}$ to another with parameter $-\sqrt{1-s^2}$ by conjugation with the unitary transformation ${\gamma_5=i \gamma_0 \gamma_1 \gamma_2 \gamma_3}$ that anti-commutes with all four Gamma matrices. Furthermore, the automaton ${\cal B}(\bold{k},s,\pm)$ is defined up to a physically-irrelevant global phase, so that we can always choose $s$ to be positive. Therefore, we can restrict ourselves to automata of the form
\begin{eqnarray}\label{solfin}
{\cal B}(\bold{k},s)=\begin{bmatrix} s \cal A(\bold{k})&  i \sqrt{1-s^2} \openone\\   i \sqrt{1-s^2} \openone& s \cal A^\dagger(\bold k)\end{bmatrix},
\end{eqnarray}
where $s$ is positive. We notice that the above automaton ${\cal B}(\bold{k},s)$ corresponds to the coupling of two Weyl automata, one right-handed and one left-handed, as discussed at the end of Section III. As in the two-dimensional case, these automata can have two different spectra depending on the choice ${\alpha=(1\pm i)/4}$ in the definition of $\cal A(\bold{k})$, leading us to two unitarily-inequivalent one-parameter families of automata. Indeed, the spectra of the automata in Eq.~(\ref{solfin}) are $\{\textrm{e}^{i \omega_{\pm}}, \textrm{e}^{-i \omega_{\pm}}\}$, each eigenvalues with multiplicity two, where
\begin{equation}\label{alpha}
\cos \omega_{\pm}=s (\cos k_x \cos k_y \cos k_z \pm \sin k_x \sin k_y \sin k_z).
\end{equation}
Here again, the $\pm$ sign corresponds to the choice ${\alpha=(1\pm i)/4}$. Finally, a convenient choice for the two one-parameter families of automata is ${\cal B}(\bold{k},s)$ and ${\cal B}^\dagger (-\bold{k},s)$ where the automaton ${\cal A}(\bold{k})$ is defined with ${\alpha=(1+i)/4}$ together with the four vectors $\bold{a}_j$ of any $B_{j-}$. In conclusion, we have found two unitary-inequivalent one-parameter families of four-dimensional homogenous, local, isotropic, and unitary automata. For small wave vectors $\bold{k}$ and large parameter $s \approx 1$, or equivalently a small positive parameter $r=\sqrt{1-s^2}<<1$, these two automata reduce to
\begin{eqnarray}
{\cal B} (\bold{k},r)=\openone + i \bold{k}.\gamma \gamma_0  + i r \gamma_0.
\end{eqnarray}
In the limit of continuous time, we obtain the Dirac equation in momentum space
\begin{eqnarray}
i\gamma_0\partial_t \Psi=\bold{k}.\gamma \Psi + r \Psi,
\end{eqnarray}
where it is tempting to interpret retrospectively the small positive parameter $r$ as the mass of a free particle with state $\Psi$ in $\mathbb C^4$.

\section{Conclusion}
In this paper, we have provided a simple derivation of the two-dimensional homogenous, local, isotropic, and unitary automata on a body-centred cubic lattice, using the notion of Gram matrix. Our derivation emphasises the link between the automata and the underlying lattice. Indeed, the transition matrices that characterise the QCA essentially are the matrix representation of the vertices of the lattice's primitive cell. We have also proven that the transition matrix at the centre of the body-centred cubic cell must be zero. We have found two unitarily-inequivalent automata that reduce to the Weyl equation in the limit of small wave vectors and continuous time. We have also briefly examined the four-dimensional case to find two unitarily-inequivalent one-parameter families of automata that both reduce to the Dirac equation in the limit of continuous time, small wave vectors, and small positive parameter $r$. More efforts will be required to solve completely the three and four-dimensional cases for a single particle. The case of interacting fields remains untouched.\\

\begin{acknowledgments}
PhR expresses his gratitude to Dr. Paolo Perinotti for useful discussions and thanks Dr. Amir Kalev as well as an anonymous referee for suggestions to improve this manuscript.\end{acknowledgments}

\end{document}